\documentclass[aps,prl,twocolumn,showpacs]{revtex4-1}
\usepackage{amsmath}
\usepackage{amssymb}
\usepackage{graphicx}

\begin{document}

\title{Quantum geometric phase in Majorana's stellar representation: \\
Mapping onto a many-body Aharonov-Bohm phase}

\author{Patrick Bruno}
\email[]{patrick.bruno@esrf.fr}
\affiliation{European Synchrotron Radiation Facility, BP 220, 38043 Grenoble Cedex, France}

\date{received 10 April 2012; accepted 26 April, 2012}

\begin{abstract}
The (Berry-Aharonov-Anandan) geometric phase acquired during a cyclic quantum evolution of finite-dimensional quantum systems is studied. It is shown that a pure quantum state in a $(2J+1)$-dimensional Hilbert space (or, equivalently, of a spin-$J$ system) can be mapped onto the partition function of a gas of independent Dirac strings moving on a sphere and subject to the Coulomb repulsion of $2J$ fixed test charges (the Majorana stars) characterizing the quantum state. The geometric phase may be viewed as the Aharonov-Bohm phase acquired by the Majorana stars as they move through the gas of Dirac strings. Expressions for the geometric connection and curvature, for the metric tensor, as well as for the multipole moments (dipole, quadrupole, etc.), are given in terms of the Majorana stars. Finally, the geometric formulation of the quantum dynamics is presented and its application to systems with exotic ordering such as spin nematics is outlined. \\
  \\ 
Phys. Rev. Lett. \textbf{108}, 240402 (2012) [http://link.aps.org/doi/10.1103/PhysRevLett.108.240402] \\
selected for a Viewpoint in \emph{Physics} [http://link.aps.org/doi/10.1103/Physics.5.65]
\end{abstract}


\maketitle

The concept of geometric phase associated with a cyclic quantum evolution \cite{berry1984, aharonov_anandan1987} has by now become a central unifying concept of quantum mechanics \cite{shapere_wilczek1989, bohm2003, chrucinski2004}. Its importance stems from the fact that its local expression, the geometric curvature, controls the quantum dynamics.

The prototype of geometric phase is that of a spin $1/2$: in this case, it can be thought of as the Aharonov-Bohm (AB) phase of a unit electrical charge moving on the Bloch sphere in the field of a Dirac monopole of unit magnetic charge located at the center of the Bloch sphere \cite{berry1984}. For a spin-$J$ system evolving within the manifold of coherent states (CSs) (i.e., quasi-classical states), this interpretation goes over straightforwardly, the electric charge being now of magnitude $2J$ \cite{berry1984}. For a spin-$J$ system in an arbitrary quantum state, except for for the specific case of a quantum evolution consisting only of global rotations (i.e., due to a, possibly time-dependent, magnetic field) \cite{bruno2004}, the geometric phase associated with an arbitrary cyclic evolution of a spin $J$ system \cite{hannay1998} does not seem to be amenable to any physically appealing AB-like interpretation, which is somehow unsatisfactory.

In the present Letter, I propose a novel theory of the geometric phase of spin systems, based upon a mapping onto a (fictitious) many-body system. From the latter emerges quite naturally a novel AB-like, physically transparent, understanding of the geometric phase of spin systems. Since any system with a finite dimensional Hilbert space of dimension $2J+1$ can be thought of as a spin-$J$ system, the present study actually holds, at least formally, for any finite quantum system.

Since any two state vectors $|\Psi_1^{(\! J)}\rangle$ and $|\Psi_2^{(\! J)}\rangle$ of the spin-$J$ Hilbert space $\mathcal{H}^{(\! J)}=\mathbb{C}^{2J+1}\! -\!\{0\}$ satisfying $|\Psi_2^{(\! J)}\rangle = c |\Psi_1^{(\! J)}\rangle$ (with $c\in \mathbb{C}, c\neq0$) yield the same expectation value for any observable, they represent the same physical state and belong to the same equivalence class ($|\Psi_2^{(\! J)}\rangle \!\! \sim \!\! |\Psi_1^{(\! J)}\rangle$); thus the manifold of physical states (projective Hilbert space) is the quotient space of equivalence classes $\mathcal{P}^{(\! J)}\equiv \mathcal{H}^{(\! J)}/\!\! \sim \ = \mathbb{C}P^{2J}$ \cite{aharonov_anandan1987}. As in Ref.~\onlinecite{hannay1998}, the approach to be used here is purely geometric and relies upon Majorana's stellar representation \cite{majorana1932} for $\mathcal{P}^{(\! J)}$. Majorana's representation is most easily understood by noticing that spin-$J$ states can be obtained as fully symmetrized states of system of $2J$ spins $1/2$ \cite{bloch_rabi1945}. This idea is at the heart of the Schwinger boson (SB) representation \cite{schwinger1952}, in which the spin-$1/2$ CS pointing along the direction $\mathbf{\hat{n}}$ of spherical angles $\theta$ and $\varphi$ is
$|\mathbf{\hat{n}}^{(1/2)}\rangle \equiv \hat{a}^\dag_\mathbf{\hat{n}} |\emptyset\rangle$,
with
$\hat{a}^\dag_\mathbf{\hat{n}} \equiv \cos\left(\frac{\theta}{2}\right) \hat{a}^\dag_\uparrow + \sin\left(\frac{\theta}{2}\right) \mathrm{e}^{\mathrm{i}\varphi}\hat{a}^\dag_\downarrow$. Let us pick $2J$ (non necessarily distinct) unit vectors
$\{\mathbf{\hat{u}}_1 ,\ldots ,\mathbf{\hat{u}}_{2J} \}\equiv \mathbf{U}$, and form the state
$| \Psi_{\mathbf{U}}^{(\! J)} \rangle \equiv \frac{1}{\sqrt{(2J)!}} \left( \prod_{i=1}^{2J} \hat{a}^\dag_{-\mathbf{\hat{u}}_i} \right) |\emptyset\rangle$.
Obviously, being a superposition of states with $2J$ SBs, such a state is a spin-$J$ state. In particular, the states
$|\mathbf{\hat{n}}^{(\! J)}\rangle \equiv \frac{1}{\sqrt{(2J)!}} (\hat{a}^\dag_\mathbf{\hat{n}})^{2J} |\emptyset\rangle$
are the spin-$J$ CSs \cite{perelomov1986}; their scalar product is given by
$
\langle \mathbf{\hat{n}}_1^{(\! J)} | \mathbf{\hat{n}}_2^{(\! J)} \rangle = \left( \frac{1+ \mathbf{\hat{n}}_1 \cdot \mathbf{\hat{n}}_2}{2} \right)^{J} \mathrm{e}^{\mathrm{i} J \Sigma \left(\mathbf{\hat{z}}, \mathbf{\hat{n}}_1, \mathbf{\hat{n}}_2\right) } ,
$
where $\Sigma \left(\mathbf{\hat{z}}, \mathbf{\hat{n}}_1, \mathbf{\hat{n}}_2\right)$ is the oriented area of the spherical triangle $\left(\mathbf{\hat{z}}, \mathbf{\hat{n}}_1, \mathbf{\hat{n}}_2\right)$, and they satisfy the following resolution of unity:
$
\mathbf{1}_{J} \equiv \frac{2J+1}{4\pi}\int_{S^2} \mathrm{d}^{2}\mathbf{\hat{n}} \ |\mathbf{\hat{n}}^{(\! J)}\rangle \langle \mathbf{\hat{n}}^{(\! J)}| .
$
The rotated SB creation and annihilation operators satisfy the commutation relations: $[\hat{a}_\mathbf{\hat{n}},\hat{a}_\mathbf{\hat{n}^\prime}] = [\hat{a}^\dag_\mathbf{\hat{n}}, \hat{a}^\dag_\mathbf{\hat{n}^\prime}]=0$ and $[\hat{a}_\mathbf{\hat{n}}, \hat{a}^\dag_\mathbf{\hat{n}^\prime}] =  \langle \mathbf{\hat{n}}^{(\! 1/2)} | \mathbf{\hat{n}}^{\prime (\! 1/2)} \rangle$. Let us introduce the CS representation $\Psi_\mathbf{U}^{(\! J)}(\mathbf{\hat{n}})\equiv \langle \mathbf{\hat{n}}^{(\! J)}| \Psi_\mathbf{U}^{(\! J)}\rangle$, which is a wavefunction over the sphere $S^2$, with probability distribution
$
  Q^{(\! J)}_\mathbf{U} (\mathbf{\hat{n}}) \equiv  |\Psi_\mathbf{U}^{(\! J)}(\mathbf{\hat{n}})|^{2}$
(Husimi function). Simple algebraic manipulations yield
$\Psi_\mathbf{U}^{(\! J)}(\mathbf{\hat{n}}) = \prod_{i=1}^{2J} \left( \frac{1-\mathbf{\hat{n}}\cdot\mathbf{\hat{u}}_i}{2} \ \mathrm{e}^{\mathrm{i}\Sigma (\mathbf{\hat{z}},\mathbf{\hat{n}},-\mathbf{\hat{u}}_i)} \right)^{1/2}$. Conversely, using the decomposition in the familiar $|JM\rangle$ basis, one sees that a generic spin-$J$ state vector $|\Psi^{(\! J)}\rangle$ can be expressed as $|\Psi^{(\! J)}\rangle = P_\Psi (\hat{a}^\dag_\uparrow, \hat{a}^\dag_\downarrow) |\emptyset\rangle$, where $P_\Psi (\hat{a}^\dag_\uparrow, \hat{a}^\dag_\downarrow)$ is a homogenous polynomial of degree $2J$ of $\hat{a}^\dag_\uparrow$ and $\hat{a}^\dag_\downarrow$, which can be factorized (up to an unimportant prefactor) in the above form $|\Psi_\mathbf{U}^{(\! J)}\rangle$ with a unique multiset $\mathbf{U}$ (Fundamental Theorem of Algebra). Thus, $\mathcal{P}^{(\! J)}$ can be univocally parameterized by the constellation $\mathbf{U}$ of $2J$ Majorana stars (MSs) (zeros $\mathbf{\hat{u}}_i$ of the Husimi function), and $|\Psi_\mathbf{U}^{(\! J)}\rangle$ can be taken as a fiducial state in $\mathcal{H}^{(\! J)}$ to describe $\mathcal{P}^{(\! J)}$. Addition or removal of $n$ stars to/from a given spin-$J$ constellation generates a state of spin $(J+n/2)$ or $(J-n/2)$, respectively. While the SB formalism was inspired by Majorana's representation \cite{schwinger2000}, the underlying geometric aspects have not been fully explored so far; to carry out this task is one of the aims of the present paper.

\begin{figure}
\includegraphics[width=1.0\columnwidth]{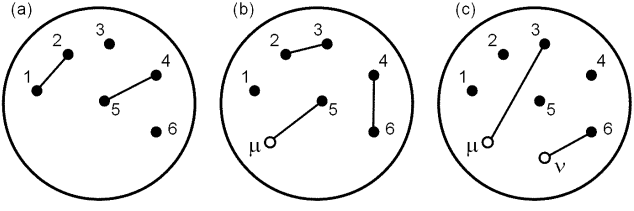}
\caption{Typical diagrams used to compute $D_\mathbf{U}^{(\! J,n)}$ (a), $D^{(\! J,n)}_{\mathbf{U}\mu}$ (b), and $D^{(\! J,n)}_{\mathbf{U}\mu\nu}$ (c). The solid dots represent the MSs (labeled from 1 to $2J$), the open dots (labeled $\mu$, $\nu$, ...) represent the auxiliary stars used to compute the multipole moments. The rules are: (i) draw all possible distinct diagrams with $n$ pairing links (each dot, solid or open, may be linked only once in a given diagram); (ii) calculate the contribution of each diagram as indicated below, and then sum over all diagrams; (iii) an unlinked solid dot yields a factor 1; (iv) an unlinked open dot yields a factor 0; (v) a link between 2 solid dots $i$ and $j$ yields a factor $d_{ij}$; (vi) a link between a solid dot $i$ and an open dot $\mu$ yields a factor $\hat{u}_{i\mu}$; (vii) a link between 2 open dots $\mu$ and $\nu$ yields a factor $-2\delta_{\mu\nu}$. The diagrams (a), (b), (c) shown here yield the contributions $d_{12}d_{45}$, $\hat{u}_{5\mu} d_{23} d_{46}$, and $\hat{u}_{3\mu} \hat{u}_{6\nu}$ to $D_\mathbf{U}^{(3,2)}$, $D^{(3,3)}_{\mathbf{U}\mu}$ and $D^{(3,2)}_{\mathbf{U}\mu\nu}$, respectively. \label{fig_diagrams}}
\end{figure}

Noticing that the two-dimensional (2D) Coulomb interaction on the sphere is $\tilde{V}(\mathbf{\hat{u}}_1,\mathbf{\hat{u}}_2)= -\ln(d_{12})$, where $2 d_{12} \equiv 2 \sin^{2}(\theta_{12}/2) = 1-\mathbf{\hat{u}}_1 \cdot\mathbf{\hat{u}}_2$ is the chordal distance between $\mathbf{\hat{u}}_1$ and $\mathbf{\hat{u}}_2$ \cite{caillol1981}, and writing
$
Q^{(\! J)}_\mathbf{U} (\mathbf{\hat{n}}) = \exp (-\tilde{\beta}\tilde{U}_\mathbf{U} (\mathbf{\hat{n}}) ) ,
$
with $\tilde{\beta}\equiv1$ and
$
\tilde{U}_\mathbf{U} (\mathbf{\hat{n}}) \equiv \sum_{i=1}^{2J} \tilde{V}(\mathbf{\hat{n}},\mathbf{\hat{u}}_i) ,
$
we can interpret the (rescaled) norm of $|\Psi_\mathbf{U}^{(\! J)}\rangle$,
\begin{eqnarray}\label{}
    \tilde{Z}(\mathbf{U}) &\equiv & \frac{\langle\Psi_\mathbf{U}^{(\! J)} |\Psi_\mathbf{U}^{(\! J)}\rangle}{2J+1} = \frac{1}{4\pi}\int_{S^2}\!\! \mathrm{d}^{2} \! \mathbf{\hat{n}} \ Q^{(\! J)}_\mathbf{U} (\mathbf{\hat{n}}) \nonumber \\
    &=& \frac{1}{4\pi}\int_{S^2}\!\! \mathrm{d}^{2} \! \mathbf{\hat{n}} \ \ \exp \left(-\tilde{\beta}\tilde{U}_\mathbf{U} (\mathbf{\hat{n}})  \right) ,
\end{eqnarray}
as the (fictitious) partition function, at inverse temperature $\tilde{\beta}\equiv1$, of a classical gas of independent particles (of density $Q_\Psi^{(\! J)} (\mathbf{\hat{n}})$) living on the sphere and interacting via the 2D spherical Coulomb repulsion with $2J$ fixed test charges located at the $\mathbf{\hat{u}}_i$'s. The corresponding fictitious free energy, $\tilde{F}(\mathbf{U}) \equiv - \tilde{\beta}^{-1} \ln \tilde{Z}(\mathbf{U})$, expresses a fictitious indirect interaction among the MSs, mediated by the gas particles in thermal equilibrium at temperature $\tilde{\beta}^{-1}$; we thus obtain a mapping of a spin-$J$ quantum state onto a $2J$-body classical system \cite{coulomb}. The partition function can be expressed in terms of the pairwise chordal distances between the MSs, $2 d_{ij}\equiv 1-\mathbf{\hat{u}}_i \cdot \mathbf{\hat{u}}_j$, as \cite{lee1988}
\begin{equation}\label{}
    \tilde{Z}(\mathbf{U} ) = \frac{1}{2J+1}\sum_{n=0}^{[J]} \left[(-1)^{n}\frac{(2J-n)!}{(2J)!}\ D_\mathbf{U}^{(\! J,n)}\right] ,
\end{equation}
with $[J]\equiv J$ (resp. $[J]\equiv J-\frac{1}{2}$) for $2J$ even (resp. $2J$ odd), and where the expression of $D_\mathbf{U}^{(\! J,n)}$ in terms of diagrams is explained in Fig.~\ref{fig_diagrams}.

Let us now show how the expectation value of the various multipole moments can be expressed in terms of the MS. They are obtained from the expectation values of the irreducible spherical tensor operators $\mathcal{\hat{Y}}_l^m (\mathbf{J})$, which in turn have the following P-representation \cite{gilmore1976}
\begin{equation}\label{}
    \mathcal{\hat{Y}}_l^m (\mathbf{J})\! = \! \frac{(2J+1+l)!}{(2J+1)! 2^l}  \frac{2J+1}{4\pi}\! \int_{S^2}\!\!\! \mathrm{d}^{2}\mathbf{\hat{n}} |\mathbf{\hat{n}}^{(\! J)}\rangle  Y_l^m(\mathbf{\hat{n}}) \langle \mathbf{\hat{n}}^{(\! J)}| .
\end{equation}
For the expectation values of the dipole moment $\hat{J}_\mu$ and the quadrupole moment $\hat{Q}_{\mu\nu} \equiv \frac{\hat{J}_\mu \hat{J}_\nu +\hat{J}_\nu \hat{J}_\mu}{2} - \frac{J(J+1)}{3} \delta_{\mu\nu}$ (where $\mu,\nu,\ldots$ label the cartesian axes), this yields
\begin{subequations}
\begin{eqnarray}
  \langle \hat{J}_\mu \rangle  &=& (J+1) \langle \hat{n}_\mu \rangle , \\
  \left\langle \hat{Q}_{\mu\nu} \right\rangle &=& (J+1)\left( J+\frac{3}{2} \right) \left( \left\langle \hat{n}_\mu \hat{n}_\nu \right\rangle -\frac{\delta_{\mu\nu}}{3} \right) ,
\end{eqnarray}
\end{subequations}
where
\begin{equation}
\langle f(\mathbf{\hat{n}}) \rangle \equiv \frac{\int_{S^2}\! \mathrm{d}^{2}\mathbf{\hat{n}}\  f(\mathbf{\hat{n}})\  Q^{(\! J)}_\mathbf{U} (\mathbf{\hat{n}}) }{\int_{S^2}\! \mathrm{d}^{2}\mathbf{\hat{n}}\  Q^{(\! J)}_\mathbf{U} (\mathbf{\hat{n}}) } .
\end{equation}
To calculate the averages $\langle \hat{n}_\mu \rangle$ and $\langle \hat{n}_\mu \hat{n}_\nu \rangle$, I remark that if we form the spin-$(J+1/2)$ state $\mathbf{U}^{\prime}$  obtained from the spin-$J$ state $\mathbf{U}$ by \textquotedblleft adding" the star $\mathbf{\hat{u}}^\prime$, and the spin-$(J+1)$ state $\mathbf{U}^{\prime\prime}$ obtained by adding one further star $\mathbf{\hat{u}}^{\prime\prime}$, the corresponding fictitious free energies are given by
\begin{subequations}
\begin{eqnarray}
  \tilde{F}(\mathbf{U}^{\prime}) &=& \tilde{F}(\mathbf{U}) + \tilde{F}(1/2) - \ln \left[ 1 - \hat{u}^\prime_\mu \langle \hat{n}_\mu \rangle \right] \\
  \tilde{F}(\mathbf{U}^{\prime\prime}) &=& \tilde{F}(\mathbf{U}) + 2\tilde{F}(1/2)   \nonumber \\
  &-& \ln \left[ 1 - (\hat{u}^\prime_\mu + \hat{u}^{\prime\prime}_\mu )\langle \hat{n}_\mu \rangle
  + \hat{u}^\prime_\mu \hat{u}^{\prime\prime}_\nu  \langle \hat{n}_\mu \hat{n}_\nu \rangle \right] ,
\end{eqnarray}
\end{subequations}
where $\tilde{F}(1/2) \equiv \ln 2$ (here and further below, Einstein's convention of summation over repeated is used, unless explicitly specified). Thus we see that we can obtain the dipole and quadrupole moments by adding 1 or 2 auxiliary stars, respectively, from the variation of the free energy as these auxiliary stars are moved around the sphere. A careful but straightforward calculation yields:
\begin{subequations}
\begin{eqnarray}
  \langle \hat{n}_\mu \rangle \! &=& \! \frac{1}{2(J \! + \! 1)} \frac{\sum_{n=1}^{[J+1 \! / 2]} (-1)^n (2J\! + \! 1 \! - \! n)! D_{\mathbf{U}\mu}^{(\! J,n)}  }{\sum_{n=0}^{[J]} (-1)^n (2J-n)! D_\mathbf{U}^{(\! J,n)}} , \\
  \langle \hat{n}_\mu \hat{n}_\nu \rangle &=&  \! \frac{1}{2(J \!\! + \! 1) (2J \! \!+ \! 3)} \frac{\sum_{n=1}^{[J+1]} (-1)^n (2J\! \! + \! 2 \! - \! n)! D_{\mathbf{U}\mu\nu}^{(\! J,n)}  }{\sum_{n=0}^{[J]} (-1)^n (2J-n)! D_\mathbf{U}^{(\! J,n)}} , \nonumber \\
  &&
\end{eqnarray}
\end{subequations}
where the expressions of $D_{\mathbf{U}\mu}^{(\! J,n)}$ and $D_{\mathbf{U}\mu\nu}^{(\! J,n)}$ in terms of diagrams are given in Fig.~\ref{fig_diagrams}. For example, the spin-1 dipole and quadrupole moments are, respectively,
\begin{subequations}
\begin{eqnarray}
  \langle \hat{J}_\mu \rangle &=& -\frac{\hat{u}_{1\mu}+\hat{u}_{2\mu}}{2-d_{12}} , \\
\!\!\!  \langle \hat{Q}_{\!\mu\nu}\! \rangle\! &=& \! \frac{1}{2\! -\! d_{12}} \! \left( \!\!\frac{\hat{u}_{1\mu}\hat{u}_{2\nu}\! +\! \hat{u}_{2\mu}\hat{u}_{1\nu}}{2} - \mathbf{\hat{u}}_1 \!\! \cdot \! \mathbf{\hat{u}}_2 \frac{\delta_{\mu\nu}}{3} \!\! \right) .
\end{eqnarray}
\end{subequations}
The extension of this procedure to higher-order multipole moments is straightforward. In turn, the method presented here allows to express, in terms of the MSs, the expectation value $H(\mathbf{U})$ of the Hamiltonian and its derivatives $\frac{\partial H(\mathbf{U})}{\partial{\mathbf{\hat{u}}_i}}$, which will be used further below to describe the quantum dynamics.

Let us now come to the geometric phase and the quantum metric. The geometric phase acquired as the systems is parallel-transported along a closed circuit $\mathcal{C}$ in $\mathcal{P}^{(\! J)}$ is given by \cite{berry1984, aharonov_anandan1987}
\begin{subequations}
\begin{eqnarray}
  \phi_{B} &=& \oint_{\mathcal{C}} \mathbf{A} \cdot \mathrm{d}\mathbf{U} \equiv \sum_{i=1}^{2J} \oint_{\mathcal{C}} \mathbf{a}_i \cdot \mathrm{d}\mathbf{\hat{u}}_i  \label{eq_Berry_a}\\
   &=& \frac{1}{2}\sum_{i,j=1}^{2J} \sum_{\alpha ,\beta =1}^2 \int_{\mathcal{S}\ (\partial\mathcal{S}=\mathcal{C})} f_{ij}^{\alpha\beta} \ \mathrm{d}\hat{u}_i^\alpha \wedge \mathrm{d}\hat{u}_j^\beta , \label{eq_Berry_b}
\end{eqnarray}
\end{subequations}
with
\begin{subequations}
\begin{eqnarray}
  a_i^\alpha &\equiv&  \frac{\mathrm{i}}{2} \left( \overrightarrow{\partial}_{\!\!\hat{u}_i^\alpha} - \overleftarrow{\partial}_{\!\!\hat{u}_i^\alpha} \right) \mathrm{ln} \langle \Psi_\mathbf{U}^{(\! J)} | \Psi_\mathbf{U}^{(\! J)} \rangle
   , \\
  f_{ij}^{\alpha\beta} &\equiv &  \partial_{\hat{u}_i^\alpha} a_j^\beta - \partial_{\hat{u}_j^\beta} a_i^\alpha = -2\ \mathrm{Im}( h_{ij}^{\alpha\beta} ) , \\
  h_{ij}^{\alpha\beta} & \equiv & \overleftarrow{\partial}_{\!\!\hat{u}_i^\alpha} \overrightarrow{\partial}_{\!\!\hat{u}_j^\beta} \mathrm{ln} \langle \Psi_\mathbf{U}^{(\! J)} | \Psi_\mathbf{U}^{(\! J)} \rangle ,
\end{eqnarray}
\end{subequations}
where $\overleftarrow{\partial}$ (resp. $\overrightarrow{\partial}$) indicates derivative of the bra $\langle \Psi_\mathbf{U}^{(\! J)} |$ (resp. ket $| \Psi_\mathbf{U}^{(\! J)} \rangle$) only. In the above equations, $\alpha,\beta =1,2$ label some spherical coordinates for the MS, with the tangent unit vectors $\mathbf{\hat{e}}_i^1$ and $\mathbf{\hat{e}}_i^2= \mathbf{\hat{u}}_i \times \mathbf{\hat{e}}_i^1$. In going from Eq.~(\ref{eq_Berry_a}) to Eq.~(\ref{eq_Berry_b}), Stokes' theorem has been used, and $\mathcal{S}$ is an oriented surface bounded by the oriented path $\mathcal{C}$. Here, $a_i^\alpha$ and $f_{ij}^{\alpha\beta}$ are, respectively, the (gauge-dependent) Berry connection and the (gauge-independent) Berry curvature tensor; they have the physical meaning of a \textquotedblleft vector potential'' and of a \textquotedblleft flux density'', respectively, in $\mathcal{P}^{(\! J)}$. The other important geometric structure is the quantum metric (Fubini-Study metric), corresponding to a distance between $|\Psi \rangle$ and $| \Phi\rangle$ defined as
$D_{\mathrm{FS}}(\Psi,\Phi )\equiv 2 \arccos \left(\frac{|\langle \Psi | \Phi\rangle|}{\langle \Psi | \Psi \rangle^{1/2} \langle\Phi | \Phi\rangle^{1/2}} \right)$, whose infinitesimal expression is
$\mathrm{d}s^2= g_{ij}^{\alpha\beta} \mathrm{d}\hat{u}_i^\alpha \mathrm{d}\hat{u}_j^\beta$, with metric tensor $g_{ij}^{\alpha\beta}= 4\  \mathrm{Re}(h_{ij}^{\alpha\beta})$ \cite{anandan_aharonov1990}.

The direct calculation of the geometric phase \cite{hannay1998} is complicated because of the need of taking care of the commutation relations among SB operators, yielding physically obscure results. This difficulty can be overcome by inserting the CS resolution of unity between bra and kets, and, after some algebraic manipulations, one obtains
$\mathbf{a}_i = \left\langle \mathbf{\tilde{a}}_i (\mathbf{\hat{n}}) \right\rangle$, where $\mathbf{\tilde{a}}_i (\mathbf{\hat{n}}) \equiv \frac{-1}{2} \left( \frac{\mathbf{\hat{z}}\times \mathbf{\hat{u}}_i }{1- \mathbf{\hat{z}}\cdot \mathbf{\hat{u}}_i} - \frac{\mathbf{\hat{n}}\times \mathbf{\hat{u}}_i }{1- \mathbf{\hat{n}}\cdot \mathbf{\hat{u}}_i}\right)$ is readily seen to be the vector potential, at $\mathbf{\hat{u}}_i$, due to a (unit flux) Dirac string entering the sphere along the $z$-axis and exiting at $\mathbf{\hat{n}}$. This means that the particles of our fictitious classical gas actually carry a Dirac string; thus the MSs are surrounded by a flux density (of total flux equal to that of a Dirac monopole of unit magnetic charge) proportional to the gas density $Q_\mathbf{U}^{(\! J)}(\mathbf{\hat{n}})$. The geometrical phase is then naturally interpreted as the AB phase acquired by the MSs as they perform a cyclic motion on the sphere. The most salient feature of this novel interpretation is the \textquotedblleft fluid'' character of the flux density, which results from the Coulomb repulsion between the flux carrying gas particles and the MSs. For a CS circuit, Berry's result \cite{berry1984} is recovered, albeit with a different AB-like interpretation. Skipping technical algebraic details, the final expression for the Berry connection is (no Einstein convention here)
\begin{subequations}
\begin{eqnarray}\label{}
    \mathbf{a}_i &=& \frac{-1}{2} \left( \frac{\mathbf{\hat{z}}\times \mathbf{\hat{u}}_i }{1- \mathbf{\hat{z}}\cdot \mathbf{\hat{u}}_i} - \left\langle \frac{\mathbf{\hat{n}} }{1- \mathbf{\hat{n}} \cdot \mathbf{\hat{u}}_i} \right\rangle \times \mathbf{\hat{u}}_i \right) \\
    &=& \frac{-1}{2} \left( \frac{\mathbf{\hat{z}}\times \mathbf{\hat{u}}_i }{1- \mathbf{\hat{z}}\cdot \mathbf{\hat{u}}_i} - \frac{ \langle \mathbf{\hat{n}}\rangle^\prime_{i} \times \mathbf{\hat{u}}_i }{1- \langle\mathbf{\hat{n}}\rangle^\prime_{i} \cdot  \mathbf{\hat{u}}_i }  \right)  \label{berry_connection_b} \\
    &=& \frac{-1}{2} \left( \frac{\mathbf{\hat{z}}\times \mathbf{\hat{u}}_i }{1- \mathbf{\hat{z}}\cdot \mathbf{\hat{u}}_i} - \partial_{\mathbf{\hat{u}}_i}\tilde{F} \times \mathbf{\hat{u}}_i \right) .
\end{eqnarray}
\end{subequations}
In Eq.~(\ref{berry_connection_b}), the notation $\langle f(\mathbf{\hat{n}}) \rangle^\prime_i$ indicates that the average is taken for the spin-$(J \! - \!1/2)$ state obtained by removing the star $\mathbf{\hat{u}}_i$ from the Majorana constellation of the spin-$J$ state $\mathbf{U}$; similarly $\langle f(\mathbf{\hat{n}}) \rangle^\prime_{ij}$, to be used further below, indicates the average taken over the spin-$(J\! -\! 1)$ obtained by removing the two stars $\mathbf{\hat{u}}_i$ and $\mathbf{\hat{u}}_j$. The first term in the above equations corresponds to the uniform flux density of a Dirac monopole, while the second one corresponds to the non-uniform part (with zero average) of the flux density.

The metric and Berry curvature and tensors are obtained in a similar manner. For the former, one gets (no Einstein convention in Eqs.~(\ref{eq_gij}--\ref{eq_gii}))
\begin{subequations}
\begin{eqnarray} \label{eq_gij}
  g_{ij}^{\alpha\beta} &=& \delta_{ij}\delta_{\alpha\beta} + \mathbf{\hat{e}}_i^\alpha \cdot \frac{\partial^2 \tilde{F}}{\partial \mathbf{\hat{u}}_i \otimes \partial \mathbf{\hat{u}}_j} \cdot \mathbf{\hat{e}}_j^\beta \nonumber \\
  &+& \mathcal{J}\mathbf{\hat{e}}_i^\alpha \cdot \frac{\partial^2 \tilde{F}}{\partial \mathbf{\hat{u}}_i \otimes \partial \mathbf{\hat{u}}_j} \cdot \mathcal{J}\mathbf{\hat{e}}_j^\beta , \\
  &=& \mathbf{\hat{e}}_i^\alpha \cdot \overline{\overline{\mathbf{d}}}_{ij} \cdot \mathbf{\hat{e}}_j^\beta +  \mathcal{J}\mathbf{\hat{e}}_i^\alpha  \cdot \overline{\overline{\mathbf{d}}}_{ij} \cdot \mathcal{J}\mathbf{\hat{e}}_j^\beta ,
\end{eqnarray}
with $\mathcal{J}\mathbf{\hat{e}}_i^1 \equiv \mathbf{\hat{e}}_i^2$, $\mathcal{J}\mathbf{\hat{e}}_i^2 \equiv - \mathbf{\hat{e}}_i^1$, and
\begin{eqnarray}
  \overline{\overline{\mathbf{d}}}_{ij} &\equiv& \left\langle \frac{\mathbf{\hat{n}}}{1- \mathbf{\hat{u}}_i \cdot \mathbf{\hat{n}}} \otimes \frac{\mathbf{\hat{n}}}{1- \mathbf{\hat{u}}_j \cdot \mathbf{\hat{n}}} \right\rangle \nonumber \\
  &-& \left\langle \frac{\mathbf{\hat{n}}}{1- \mathbf{\hat{u}}_i \cdot \mathbf{\hat{n}}} \right\rangle \otimes \left\langle \frac{\mathbf{\hat{n}}}{1- \mathbf{\hat{u}}_j \cdot \mathbf{\hat{n}}} \right\rangle .
\end{eqnarray}
\end{subequations}
For $i\neq j$, this yields
\begin{eqnarray}\label{}
    \overline{\overline{\mathbf{d}}}_{ij} &=& \frac{\langle \mathbf{\hat{n}}\otimes \mathbf{\hat{n}} \rangle^\prime_{ij}  }{1- ( \mathbf{\hat{u}}_i + \mathbf{\hat{u}}_j ) \cdot \langle \mathbf{\hat{n}}\rangle^\prime_{ij}  + \mathbf{\hat{u}}_i \cdot \langle \mathbf{\hat{n}}\otimes \mathbf{\hat{n}} \rangle^\prime_{ij} \cdot \mathbf{\hat{u}}_j  } \nonumber \\
    &-& \frac{\langle \mathbf{\hat{n}} \rangle^\prime_{i} }{1 - \langle \mathbf{\hat{n}} \rangle^\prime_{i} \cdot \mathbf{\hat{u}}_i  } \otimes \frac{\langle \mathbf{\hat{n}} \rangle^\prime_{j} }{1 - \langle \mathbf{\hat{n}} \rangle^\prime_{j} \cdot \mathbf{\hat{u}}_j  } ,
\end{eqnarray}
whereas for $i=j$, one gets
\begin{equation}\label{eq_gii}
g_{ii}^{\alpha\beta} = \delta_{\alpha\beta} \frac{ 1 - \left( \langle \mathbf{\hat{n}} \rangle^\prime_{i} \right)^2 }{\left(1 - \langle \mathbf{\hat{n}} \rangle^\prime_{i} \cdot \mathbf{\hat{u}}_i \right)^2 } .
\end{equation}
The Berry curvature tensor is obtained from the metric tensor from the following identities
\begin{subequations}
\begin{eqnarray}
 -2 f_{ij}^{12} &=&  2 f_{ij}^{21} = g_{ij}^{11} = g_{ij}^{22} , \\
  2 f_{ij}^{11} &=&  2 f_{ij}^{22} = g_{ij}^{12} = -g_{ij}^{21} .
\end{eqnarray}
\end{subequations}
Mathematically, this follows from the K\"{a}hlerian nature of the projective Hilbert space \cite{chrucinski2004}.

To describe the quantum dynamics, I write down Schr\"{o}dinger's equation for the state vector $|\psi (t) \rangle \equiv \mathrm{e}^{\mathrm{i}\varphi (t)} \frac{|\Psi_{\mathbf{U}(t)} \rangle}{\langle \Psi_{\mathbf{U}(t)} |\Psi_{\mathbf{U}(t)} \rangle^{1/2}}$, which yields $\dot{\varphi}  = - H(\mathbf{U})+ \mathbf{A}\cdot \dot{\mathbf{U}}$ (setting $\hbar \equiv 1$); the latter result is nothing but as the infinitesimal version of the Aharonov-Anandan decomposition of the total phase variation into the dynamical and geometric terms \cite{aharonov_anandan1987}. It is not gauge invariant with respect to a change of the phase choice for the fiducial states $| \Psi_\mathbf{U}\rangle$. In order to obtain a gauge invariant equation of motion, we use the fact that, due to the unitarity of Hamiltonian evolution, the relative phase of any two (non-orthogonal) states, defined as $\varphi_{12}(t) \equiv \arg \langle \psi_1 (t) | \psi_2(t) \rangle$, is time-independent. Doing this for states at $\mathbf{U}$ and $\mathbf{U}+\delta\mathbf{U}$, one finally obtains
\begin{equation}\label{eq_dynamics}
  f_{ij}^{\alpha\beta} \partial_t \hat{u}_j^\beta = \partial_{\hat{u}_i^\alpha}H(\mathbf{U}) .
\end{equation}
The above equation has the form of the classical equation of motion of a system of $2J$ coupled particles evolving on a spherical phase space, with symplectic form given by $\frac{1}{2} f_{ij}^{\alpha\beta} \mathrm{d}\hat{u}_i^\alpha \wedge \mathrm{d}\hat{u}_j^\beta$, and a Hamilton function given by $H(\mathbf{U})$ \cite{arnold1978}. An alternative (equivalent) formulation of the spin dynamics in terms of the Majorana stars has been given earlier by Leb{\oe}uf \cite{leboeuf1991}; however, the symplectic-Hamiltonian nature of the dynamics is displayed more transparently in the present formulation. I note that the $2J$ particles are coupled to each other, not only dynamically through the Hamilton function, but also kinematically via the symplectic form. Eq.~(\ref{eq_dynamics}) is the quantum mechanical counterpart of the Landau-Lifshitz equation for spin dynamics \cite{landau_lifshitz1935}, which we would recover if we would restrict our description to (quasi-classical) CSs. For quantum spin systems, such as molecular magnets \cite{gatteschi2006}, the latter is clearly inadequate, and a fully quantum description as given in Eq.~(\ref{eq_dynamics}) is necessary.

Finally, I briefly address the question of systems of interacting spins in magnetically ordered systems. The usual treatment, spin-wave theory, amounts to use a variational wave-function given as a tensorial product of CS with site-dependent unit vectors $\mathbf{\hat{n}}(\mathbf{R}_i,t)$, and solve the linearized coupled Landau-Lifshitz equations. This approach is clearly not adequate for systems with exotic (e.g., quadrupole or higher-multipole) ordering such as spin nematics \cite{chen_levy1971}, and the effective field theory proposed for spin nematics \cite{ivanov_kolezhuk2003} applies only to spin-1 systems and cannot be extended to systems with spin $J>1$. This clearly calls for a more general theory of spin nematics. I argue here that the most natural description of such systems is the geometric description offered by the Majorana representation. It consists in setting up a quantum field theory based upon Majorana's constellations instead of the quasi-classical CSs; one thus obtains a path-integral theory for the $2J$ coupled $O(3)$ fields $\mathbf{\hat{u}}_i$. This theory will be developed in detail in a forthcoming paper.

I gratefully acknowledge helpful discussions and/or correspondence with Robert Whitney, Efim Kats, and John Hannay.
\begin{acknowledgments}
\end{acknowledgments}

\end{document}